\documentclass[twocolumn,aps,pra]{revtex4}

\usepackage{amsmath, amssymb, amsfonts}
\usepackage{bm, mathrsfs, dsfont,bbold}
\usepackage{braket}
\usepackage{graphicx}
\usepackage{multirow}
\usepackage{color}
\usepackage{array}

\newcommand{\ketbra}[2]{\ket{#1}\!\!\bra{#2}}

\graphicspath{{./}{./pics/}}

\begin{document}

\title{Absorption-based Quantum Communication with NV centres}

\author{Burkhard Scharfenberger$^1$, Hideo Kosaka$^2$, William J. Munro$^{3,1}$ and Kae Nemoto$^1$}

\affiliation{$^1$ National Institute of Informatics, 2-1-2 Hitotsubashi, Chiyoda-ku, Tokyo 101-8430, Japan}
\affiliation{$^2$ Department of Physics, Yokohama  National University, Sogo-Kenkyuto S306, Tokiwadai, Hodogayaku, 240-8501 Yokohama, Japan}
\affiliation{$^3$ NTT Basic Research Laboratories, NTT Corporation, 3-1 Morinosato Wakamiya, Atsugi, Kanagawa 243-0198, Japan}

\date{\today}

\begin{abstract}
We propose a scheme for performing an entanglement-swapping operation within a quantum communications hub (a Bell like measurement)  using an NV- centre's $|\pm 1\rangle \leftrightarrow |A_2\rangle$ optical transition. This is based on the heralded absorption of a photon resonant with that transition. The quantum efficiency of a single photon absorption is low but can be improved by placing the NV center inside a micro cavity to boost the interaction time and further by recycling the leaked photon back into the cavity after flipping its phase and/or polarisation. Throughout this process, the NV is repeatedly monitored via a QND measurement that heralds whether or not the photon absorption has succeeded. Upon success we know a destructive Bell measurement has occurred between that photon and NV center. Given low losses and a high per-pass absorption probability, this scheme allows the total success probability to approach unity. With long electron spin coherence times possible at low temperatures, this component could be useful within a memory-based quantum repeater or relay. 
\end{abstract}

\pacs{}

\maketitle

\begin{figure*}[htb]
\includegraphics[width=0.88\textwidth]{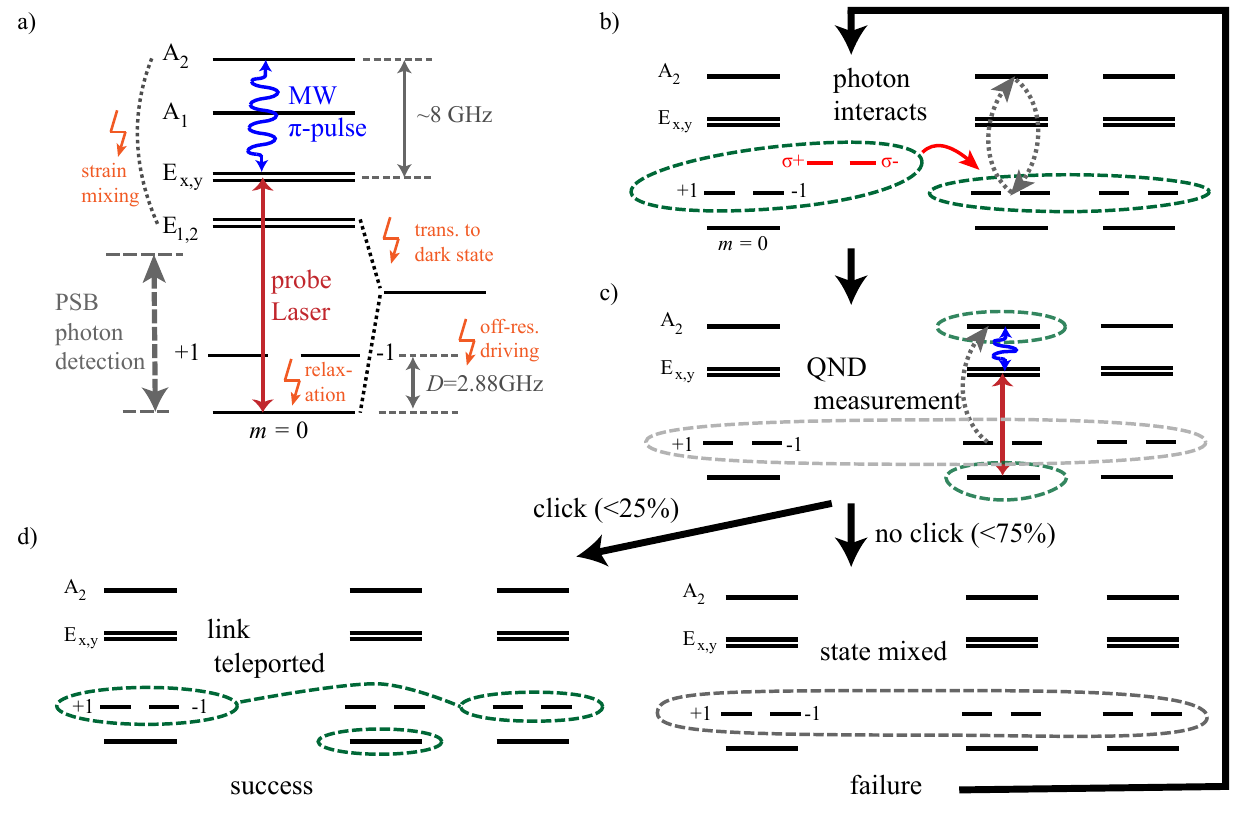}
 \caption{An illustration of the QND measurement and its use in teleportation.
 		a) QND measurement and potential error channels (orange symbols).
		    Off-resonant driving can by strongly suppressed by timing the length of the $\pi$-pulse.
 		b) a spontaneously emitted photon entangled with the left NV travels to and interacts with the NV in the middle,
		    which is entangled with a third NV, causing coherent oscillations between $\ket{\pm 1}$ and $\ket{A_2}$.
		c) when the amplitude of  $\ket{A_2}$ is maximal (within the photon residence time $\tau_\text{int}$) 
		    the QND measurement is applied 
		d) receiving PSB photons ('click') heralds projection onto $\ket{A_2}$ and teleportation of the
		    entangled link. If no signal is seen within some time window, the state has become completely 
		    mixed (assuming the photon actually interacted)
		 }
 \label{fig:teleport}
 \end{figure*}

\section{Introduction}


Quantum communication \cite{BB84,Bennett96,enk98,Gisin2002,Gisin2007} is a resource that will be required in the development of tomorrow's quantum internet \cite{Kimble2008}  whether it be to share quantum enabled information over short, medium or long ranges. It will enable a multitude of tasks ranging from quantum key distribution (QKD) \cite{Gisin2002}, device independent QKD \cite{Vazirani-14prl140501} to distributed quantum computation and sensing \cite{cirac1999,meter2009,Giovannetti2004}. Especially over long ranges (hundreds or thousands of kilometres) one will require shared entanglement to enable this, which will most likely require the use of quantum repeaters \cite{bennett1993,Briegel1998,Sangouard2011}.  These take the form of a chain of nodes able to perform two basic functions: the first being to store already established entangled links between nodes and the second to  merge two links into a new, longer one. Many repeater schemes using various approaches to provide this functionality and various physical systems to implement the nodes of such a repeater have been put forward over the years \cite{Bennett96,enk98,Briegel1998,Sangouard2011,dur98,duan01,loock06,chen07a,sheng08,goebel08,simon07,tittel08,sangouard09,pan01,dur07,munro10,munro12}, including negatively charged nitrogen vacancy center in diamond (NV-)~\cite{Childress2006,Stephens-13pra052333}.

The NV- center  is made up of a vacant carbon position in a diamond lattice adjacent to a substitutional 
nitrogen atom \cite{Davies1976,Harley1984,VanOort1988,Doherty2013} has many desirable properties \cite{Childress2005,Childress2006,Dutt2007,Jiang2007,YJG12,Maurer2012,Togan2010,Jelezko2004}, including long coherence times \cite{Maurer2012} and 
good controllability of both electronic and nuclear spins at different temperatures 
~\cite{Kennedy-03apl4190,Jelezko2004,Jelezko-04prl130501,Harrison-04jlum245,
Bala-09nm383,Fuchs-10np668,wang-12prb155204,Dutt2007,Neumann-08s1326,Neumann-10s542544,
Robledo-11n574,Maurer2012,Waldherr-13n204,Taminiau-13nn171}. 
The experimentally accessible states are the ground state manifold (GSM) and the excited state 
manifold (ESM), both of which are spin triplets but the former is composed of orbital angular momentum
singlets (A$_2$) while the states of latter form orbital doublets (E) under the action of the defect's 
symmetry group C$_\text{3,V}$ \cite{Doherty2013}. The two sets of states are separated by an optical transition with 
zero-phonon-line wavelength of 637nm \cite{Togan2010} while a set of optically inaccessible intermediate spin-singlet 
states can be reached from the ESM via non-radiative decay processes \cite{Doherty2013}. 

NV- centres have already been used as emitter of two different kinds of entangled photon qubits: 
polarisation~\cite{Togan-10n09256} and time-bin~\cite{Bernien-13n12016}.  The former makes 
use of the special Bell-state like form of the four $m_S=\pm 1$ zero field ESM states. 
In particular the $\ket{A_2}$ state does not couple to the intermediate singlet states. Since in a dipole interaction the z-component of
total the angular momentum must change by $\pm 1$, spontaneous emission out of $\ket{A_2}$ creates
a photon which is polarisation-entangled with the two degenerate $m_S=\pm 1$ states of the GSM \cite{Togan2010}.
It has been recently demonstrated experimentally, that the time reversal also holds: absorption of
a photon acts like a projection onto a joint photon-vacancy spin Bell-state~\cite{Kosaka-15prl053603}. 
In principle this could allow for the teleportation of quantum information from an incoming photon 
to any qubit the absorbing NV- is entangled with.

Our proposal is to use this feature of the NV center as a core element of a quantum communications network. A node in the 
network  can perform an  entanglement swap operation upon heralded absorption of a photon carrying an entangling 
link to a remote NV-. The heralding (QND type) measurement is implemented via detecting the phonon-side band photons
of a classical laser pulse tuned to the transition between the $m_S =0$ state of GSM and ESM after 
a microwave $\pi$-pulse resonant with the $\ket{A_2}$ to $\ket{E_{x|y}}$ ($m_S=0$) transition.
To enhance interaction, the NV- would be placed inside a micro cavity but even in this case the 
single-pass absorption quantum efficiency would be limited to 25\%. We can however increase this further by 
adding a fibre-optic loop to re-cycle leaked photons back into the cavity after
potentially flipping its phase and/or polarisation.


This paper is structured as follows: in section II we describe the main idea of 
the entanglement swapping scheme, detailing two possible approaches to increase 
quantum efficiency beyond 25\%. We then proceed in In section III to present a model of the scheme  
as a discrete-time Quantum Markov process including estimations for the various parameters appearing.  
We show the results of numerical simulations using this model for different parameter regimes, allowing us to make quantitative  
statements about the expected performance of our scheme taking into account 
losses and errors. In section IV we briefly discuss how this heralded Bell measurements can be used in quantum repeaters and relays. 
We summarise our findings in section V.
 
\section{Teleportation scheme}
Here we will describe in detail the three main ingredients of our scheme,
the entangling absorption operation, the QND measurement and photon recycling.
 
\emph{Entangling absorption. ---}
At zero magnetic field and zero strain the ESM A$_2$  state has the form 
  $\ket{\text{A}_2}=(\ket{E_+,-1} + \ket{E_-, +1})/\sqrt{2}$
where $E_\pm$ and $\pm 1$ denote the (collective) orbital angular momentum and spin 
of the NV$^-$ vacancy respectively. This is a $\psi_+$ Bell state between the orbital and spin
degree of freedoms. From this $\ket{\text{A}_2}$ state there are dipole-allowed optical transitions to
the GSM $m_S=\pm 1$ states which due to angular momentum conservation are dependant 
on the photon polarisation. 

This property of the A$_2$ state was exploited in~\cite{Togan-10n09256} to 
generate polarisation-entangled photons by spontaneous emission out of
the $\ket{A_2}$ state. However, this `entangling emission', can also be time-reversed into an `entangling absorption'
of an incoming photon. As was demonstrated in a recent experiment absorption into $\ket{A_2}$ is 
equivalent to a Bell-measurement on the joint photon-vacancy electronic 
spin system of the form~\cite{Kosaka-15prl053603}
\begin{equation}
  M_{\psi_+} \! = \frac{1}{2}\left(\ket{+1,\sigma_-}\! + \! \ket{-1,\sigma_+}\right)\!\left(\bra{+1,\sigma_-}\! + \! \bra{-1,\sigma_+}\right)
\end{equation}
where the first quantum number refers to the vacancy spin and the second
to the circular photon polarisation states $\sigma_\pm$.
This implies, that if the absorbing NV centre starts out maximally entangled 
with some other qubit, whatever information the photon was carrying will be teleported to 
that qubit. This is basic process which enables our proposed quantum repeaters and relays.
 
 \emph{QND measurement. ---}
 It is important to realise, however, that the teleportation does not happen until NV centre
 is detected in the $\ket{A_2}$, i.e. until the state is projected onto $\ket{A_2}$.
 Without this, the interaction with the photon will simply cause coherent flip-flops between the
 NV's initial state (locally a completely mixed state of $m_S=\pm 1$) and $\ket{A_2}$ at a 
 frequency determined by the interaction strength.
 We propose to implement this projective QND measurement via a 
 microwave $\pi$-pulse resonant with the transition from $\ket{A_2}$ to the ESM $m_S=0$ states
 $\ket{E_{x|y}}$ and simultaneous irradiation with a probe laser beam tuned to the transition
 between the $m_S = 0$ states in the GSM and ESM ($\approx$637nm). Detection of phonon-side-band
 (PSB) photons would then herald, that the NV indeed was in $\ket{A_2}$, while seeing no
 signal can with high confidence be interpreted to mean the system is still in its initial 
 state.
A schematic of the QND measurement and how it is employed as part of the teleportation
scheme is depicted in figure~\ref{fig:teleport}. In addition to the ideal process, 
~\ref{fig:teleport}a also shows possible error channels which limit the fidelity of the QND
measurement, causing either false positive (off-resonant driving, relaxation) or 
indeterminate errors (strain mixing plus transition to dark state). 
We should point out, that since the measurement is projective, the wave function changes
not only in the desired case of seeing a heralding signal. Rather, also in the case where no
signal is received a change has occurred, given the photon actually interacted, 
which can be described by the operator $1-M_{\psi_+}$ with $M_{\psi_+}$ the entangling 
measurement operator from above.
The application of the QND measurement $\pi$ pulse would be timed to coincide with
the time of maximum probability to find the NV in $\ket{A_2}$. This is going yield
quite low probabilities of absorption for a free photon pass. Therefore, we envision 
to place the NV inside a micro cavity which both enhances interaction strength and 
interaction time (residence time) between photon and NV. 
The cavity would have to be carefully designed to fulfil this purpose while at the same
time allowing the photon to enter without being reflected.
 
\begin{figure}
\includegraphics[width=0.375\textwidth]{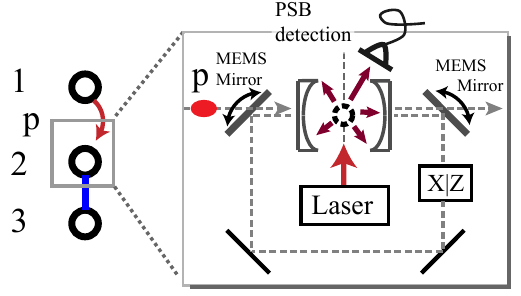}
   \caption{    Sketch of the recycling loop: 
		      a micro cavity enhances NV photon interaction in order to get a higher 
		      absorption probability per pass. Since manipulating the
		      NV is not possible while the QND measurement is on
		      the photon state. This is done outside the cavity by either actively (approach A)
		      or passively (approach B) flipping the photon's phase/polarisation
		      before sending it back into the cavity.
   		}
   \label{fig:recycling}
\end{figure}

\emph{Mismatch problem and recycling loop. ---}
However, even with perfect absorption the incoming photon will only 
be absorbed at most 25\% of the time if no further measures are taken.
This is easily understood by looking at the full initial state 
$\ket{\psi_0}=\frac{1}{2}(\ket{+1,\sigma_-} + \ket{-1,\sigma_+})_{1\text{p}}(\ket{+1,+1}+\ket{-1,-1})_{23}$,
where '2' labels the NV the photon p interacts with and 1 and 3 are remote 
qubits, which in the following we will always assume to be other NV centres. 
Rewriting $\ket{\psi_0}$ in the Bell-basis between qubits 1 and 3 we find
\begin{equation}
  \begin{aligned}
  \ket{\psi_0} = \frac{1}{2} & \left(\ket{\phi_+}_{13}\ket{\psi_+}_{2p} + \ket{\phi_-}_{13}\ket{\psi_-}_{2p} \right. \\
    		       &  \left. + \ket{\psi_+}_{13}\ket{\phi_+}_{2p} + \ket{\psi_-}_{13}\ket{\phi_-}_{2p}   \right) .
  \end{aligned}
  \label{eq:psi0}
\end{equation}
 where the $\ket{\phi_\pm}$ and $\ket{\psi_\pm}$ denote the even and odd parity Bell-states
 respectively.
 Only the $\ket{\psi_-}_\text{2p}$-term allows a dipole transition to $\ket{A_2}_2$, the dipole operator
 matrix elements $\braket{x | \vec{E}\cdot\vec{r} |A_2}$ for the three other states ($x=\psi_+, \phi_\pm$) 
 are zero. Their symmetry matches the other three $m_S=\pm 1$ states of the ESM,
 which are however detuned by at least about 3GHz resulting in a relative
 transition probability ratio of at worst $p_{\text{abs, A}_1},/p_{\text{abs, A}_2}<10^{-4}$.
 If we want a high overall success probability we therefore need to somehow turn the
 other $2p$ Bell states into $\ket{\psi_-}_\text{2p}$ by applying some operation to either 
 the spin or the photon. 
 As ilustrated in Figure~\ref{fig:recycling}, our idea to raise the quantum efficiency closer
 to 1 is then to let the photon interact with the NV center inside a cavity with residence time 
 $\tau_\text{int}$ determined by the cavity Q-factor, and, via the QND measurement, check 
 whether the NV has transitioned into $\ket{A_2}$. 
 If the heralding signal is seen the link is teleported with very high 
 probability reduced only by the probability for a false positive, which we estimate to be
 low (see appendix).
 If no heralding signal is seen within time $\tau_\text{int}$ we can assume, also with high
 probability, that no absorption occurred and the photon left the cavity after time $\tau_\text{int}$. 
 Outside the cavity, the photon would be caught by a fibre-optic loop  which serves route it back
 into the cavity. Furthermore the loop will contain an integrated(active or passive) switch
 allowing us to flip the photons's polarisation, phase or both.
 if this 'recycling'  process takes time $\tau_\text{out}$, the total (average) time per cycle is 
 $\tau= \tau_\text{int} + \tau_\text{out}$. 
 We would repeat this procedure a predetermined number of times $L$, before finally
 abandoning the teleportation attempt.
 In the ideal case of perfect absorption per pass we need to re-cycle the photon 
 only three times, corresponding to $L=4$, to get maximal probability of success. 
 However, in practice absorption cannot be perfect even with a cavity, and we need 
 to recycle multiple times per polarisation/phase setting. 
 
 \subsection{Two approaches}
 Within the framework described above there are, at least, two ways 
 of solving the mismatch problem: the technically simpler one, which we will call
 approach A, is to flip only either 
 the phase or polarisation on each cycle. After an even number $L=2k$ of cycles
 without absorption we measure both the photon and the vacancy spin in the XX 
 or ZZ basis, depending on whether we chose to flip phase or polarisation.
 Assuming we flip the phase, even parity outcomes ('++' or '- -') correspond to a
 the Bell state $\ket{\psi_+}_{13}$ and odd parity ('+-' or '-+' ) to $\ket{\psi_-}_{13}$.
 As we will see later, while technically simpler to implement, this scheme requires
 high per-pass absorption $p_\text{abs}$ to achieve good fidelities for the teleported link
 as well as highly efficient single photon detectors.
 
 In another approach, here dubbed B, without these limitations both phase and 
 polarisation are flipped periodically after $l_z$ and $l_x$ rounds respectively.
 Choosing $L = 2 l_x = 4 l_z$ gives each of the four possible Bell-states a (roughly) 
 equal chance of being heralded by the QND measurement, making a final XX or ZZ 
 measurement unnecessary. With this approach, link fidelities are almost independent of $p_\text{abs}$ 
 and decrease with the number of cycles L due to dephasing of the electronic spin, dark counts 
 (false positive signals) and transitions to ESM states other than $A_2$.  It can thus be used in 
 the low per-pass absorption probability regime but has the downsides of greater technical 
 complexity as well as a lower total success 
 probability $p_\text{success}$.

 
\section{Modelling}
 In this section we present a discrete time Quantum Markov model that we used to obtain 
 quantitative results about the performance of our recycling-loop teleportation scheme when
 applied to a situation as shown in figure~\ref{fig:recycling} taking into account real world 
 imperfections. 
 We restricted the description to the 32 dimensional effective Hilbert space spanned by 
 $\{\ket{\phi_\pm}, \ket{\psi_\pm} \}_{13} \otimes \{\ket{\phi_\pm}, \ket{\psi_\pm}; \ket{A_2}, \ket{A_1}, \ket{\pm 1} \}_{2[p]}$.
 Starting in the state $\rho_0 = \ketbra{\psi_0}{\psi_0}$ with $\psi_0$ as defined in 
 the previous section we loop through the following steps $L$ times:
 \begin{itemize}
   \item[1.] absorption: with probability p$_\text{Abs} \braket{\psi_-|\rho|\psi_-}_{2p}$ the photon will be
   		absorbed and the NV transitions to the A$_2$ state
   \item[2.] QND measurement: the NV is measured and with probability $p_\text{click} = p_\text{QND}\braket{\text{A}_2|\rho|\text{A}_2 }_2+p_\text{Dark}$
   		absorption is heralded, and with prob. $1-p_\text{click}$ it is not. In either case the state $\rho$ 
		is updated accordingly 
  \item[3.]  photon loss: with probability $p_\text{loss}$, the photon is lost during the recycling process
  \item[4.]  dephasing: a dephasing operation with $\eta_2=\exp(-\tau^2/T_2^2)$ is applied to all involved NVs 
  		(1,2 and 3) in the $(\ket{+1}\pm \ket{-1})/\sqrt{2}$ basis where $\tau$ is the time per cycle
  \item[5.]  flipping: every $l_z$-th ($l_x$-th) loop a phase (polarisation) flip is 
  		applied to the photon
 \end{itemize}
 Since imperfections in the phase and polarisation flip operations do likely not play a major role, we 
 did not include them in this model.
 Furthermore, as is readily apparent, the success probability depends very strongly on the photon actually 
 arriving at and entering into the our cavity and recycling loop structure. 
 In fact, the chance that the photon is lost are quite high for fibre bound communications over many 
 kilometres. However, since here we are interested only in the performance of our scheme as a component
 of a network we do not take this initial loss probability into account.
 
 \subsection{Parameter Estimation}
 Thus we have a total of five parameters relevant to the performance during each cycle
 ($p_\text{abs}$, $p_\text{QND}$, $p_\text{Dark}$,
 $p_\text{loss}$ and $\eta_2$) and, depending on the approach (A or B), one or three more: 
 total number of rounds $L$ (A and B), polarisation-flip period $l_x$ and phase-flip period $l_z$ (only B). 

 For both approaches we have a fraction of false-negatives approximately $1-p_\text{QND}$ of 
 cases in which we discard a correctly teleported link (see appendix~\ref{app:errors}), which is 
 then erased when we try again. 
 But since the QND measurement has high fidelity this is not likely to be a limiting factor. 
 
 
 This brings us to the question: what are realistic values for the parameters in the model? 
 As we already stated, the QND measurement is likely to be high and the value
 of $p_\text{QND} = 99$\% we used in our simulations is likely to be conservative. 
 For $p_\text{Dark}$ we use an equally conservative 0.01\%. 
 While a photon loss of 1dB per element is usually regarded as good, here we cannot tolerate
 more than $\approx$1dB for the total structure. We investigated values from 0 to 0.5dB but all
 explicit references to performance estimates are for the challenging but potentially achievable value 
 of $p_\text{loss}=0.3$dB.
 Furthermore, assuming a dephasing time of $T_2=100\mu\text{s}$ and total time per round of 
 $\tau=200$ns we find that the dephasing error is about $1-\eta_2=4\times 10^{-6}$ and thus quite small. 
 
 This leaves open the probability of absorption (per pass) $p_\text{abs}$ and the number of 
 rounds $L$ which in the next section we will treat as the variable of our analysis. 

   \begin{figure*}
   \hspace{.09\textwidth}{\footnotesize Approach A} \hspace{0.275\textwidth}{\footnotesize Approach B}
   \includegraphics[width=0.975\textwidth]{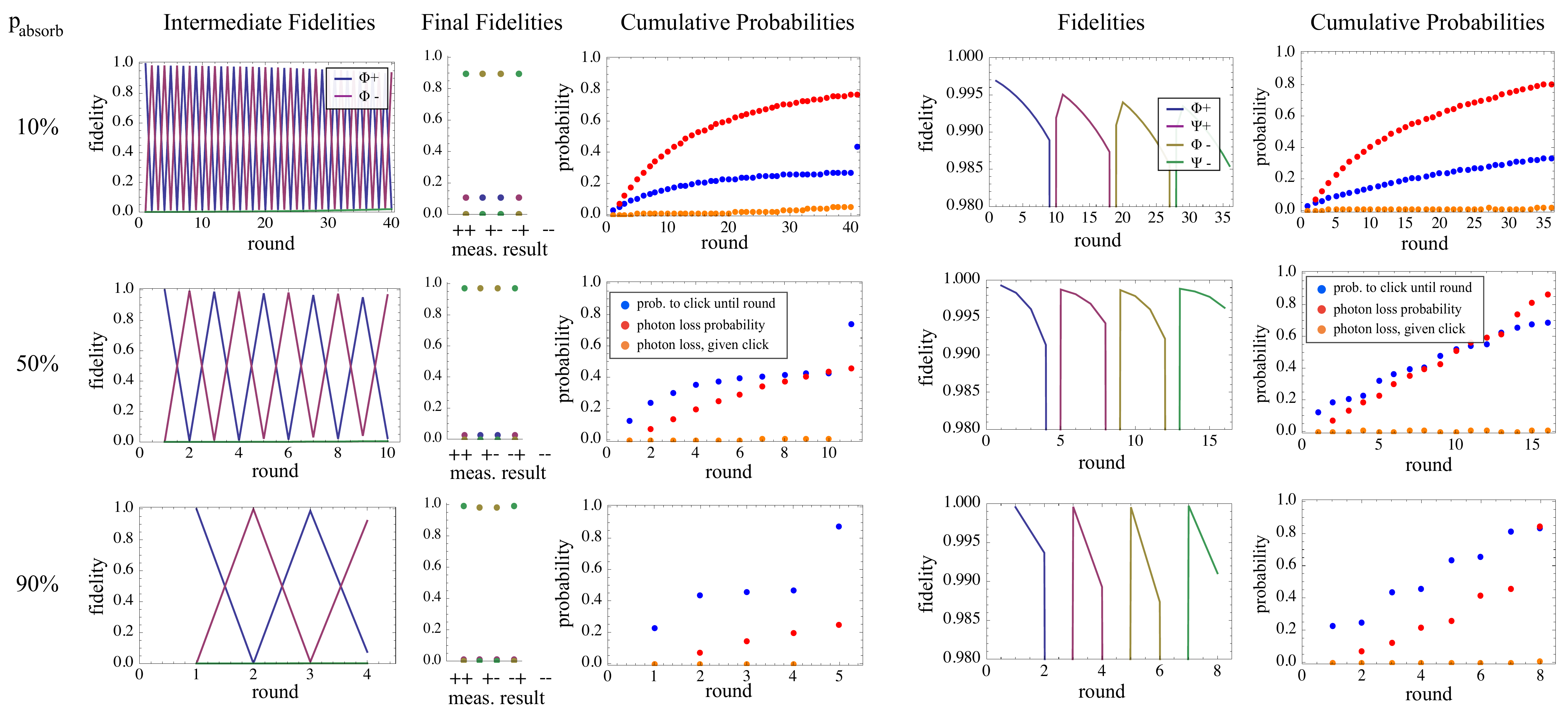}
    \caption{Approach A and B link fidelity and success probability (=cumulative probability of seeing a click) 
    		vs number of rounds for low, medium and high absorption probability 
		and optimal $l_z$, $l_x$ (only B) and $L$ (both).
		As can be seen the link fidelity declines with time mostly due to the effect of 
		off-resonant transitions to $\ket{A_1}$.
    		}
      \label{fig:scheme3}
 \end{figure*}
 
 \subsection{Analysis}
 With parameters as determined in the previous section, we tested multiple combinations
 of $p_\text{absorption}$ and $L$ as well as $l_x$ and $l_z$ (only approach B).
 In the latter case the best fidelities are naturally obtained when attempts are equidistributed over
 all four possible Bell states between NV centres 1 and 3.
 
 The resulting success probabilities (cumulative probability to see a heralding signal) and 
 average link fidelities for low, medium and high absorption probability and optimal $L$ (and $L=2l_x = 4l_z$ in
 case of approach B) are shown in figure~\ref{fig:scheme3} as
 a function of recycling round.
 
 We find, perhaps unsurprisingly, that the probability of absorption is indeed of critical importance. 
 However the simulations also show that there are clearly diminishing returns: the two figures of merit 
 total success probability and average link fidelity increase much more from the low to medium 
 $p_\text{abs}$-regime than from the medium to high regime. 
 Consequently it seems advisable to try to increase $p_\text{abs}$ to, if possible, 
 at least around 50\%, but it might not be worthwhile pushing far beyond this. 
 At this point it is unclear, how far $p_\text{abs}$ can be improved by use of a cavity, but
 should it prove difficult to reach or go much beyond 50\% technical efforts should rather 
 be focused on minimising the per cycle losses $p_\text{loss}$.
 
\begin{figure*}
   \hspace{-0.4\textwidth} a) \hspace{0.4\textwidth} b)\\
      \includegraphics[width=0.42\textwidth]{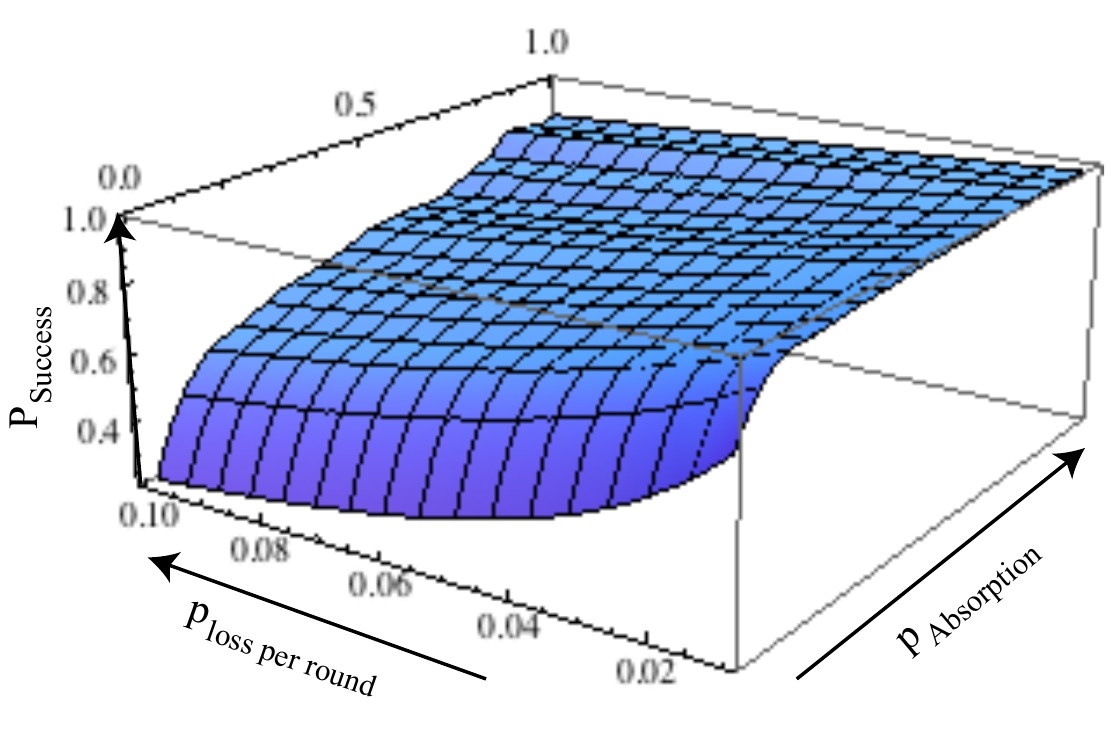}\hspace{0.02\textwidth}\includegraphics[width=0.4\textwidth]{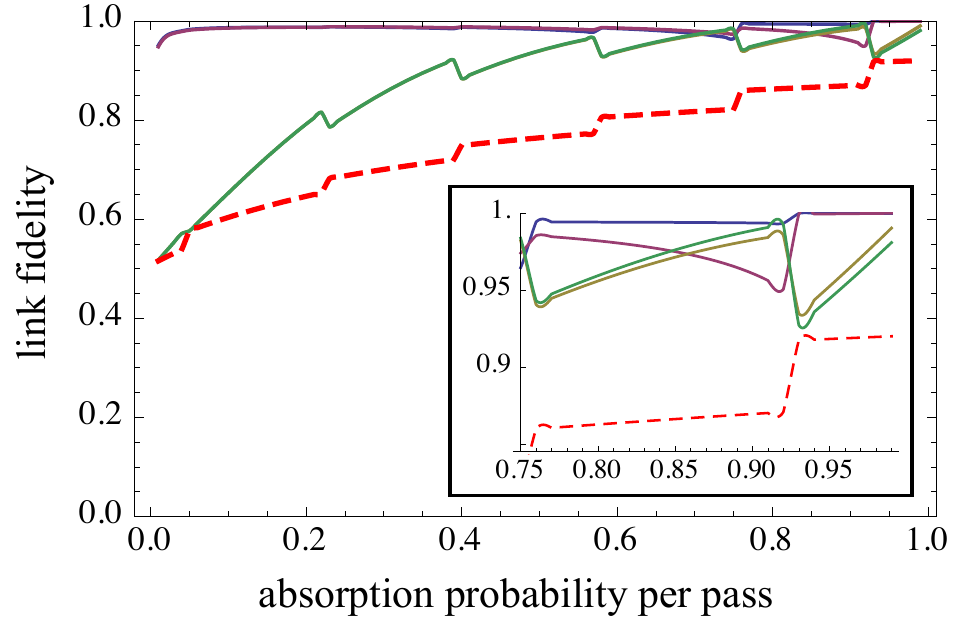}\\
  \hspace{-0.4\textwidth} c) \hspace{0.4\textwidth} d)\\
   \includegraphics[width=0.42\textwidth]{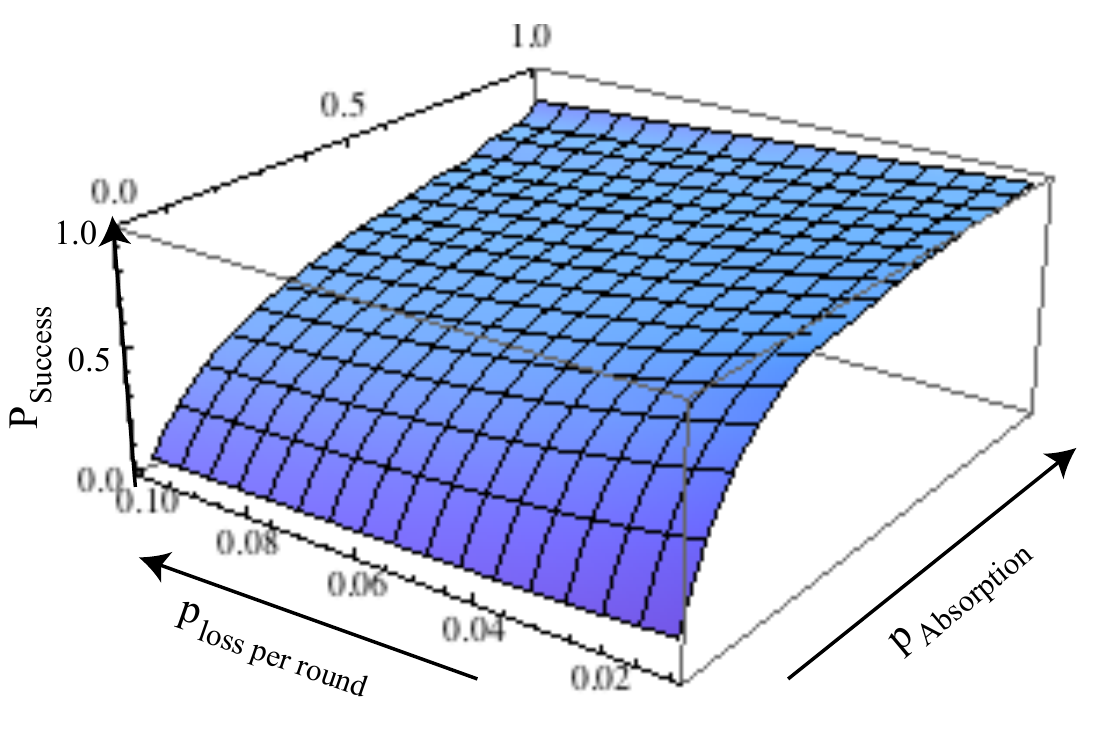}\hspace{0.02\textwidth}\includegraphics[width=0.4\textwidth]{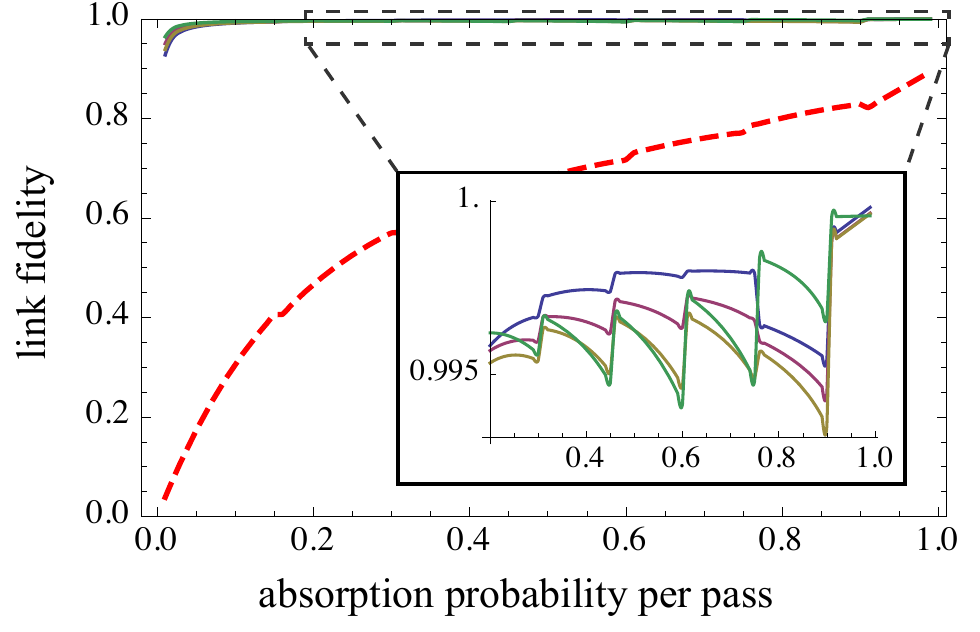}
    \caption{Success probabilities and fidelities for Approaches A (top) and B (bottom)
    		 left: total success probability plotted over probability of absorption and loss per pass.
    		 right: Link fidelities for the four possible output Bell states plotted vs. per-pass probability
		 of absorption for a per-pass loss probability of 6.6\% (corresponding to 0.3dB). 
		 In b and d the insets show a zoom in on the region $p_\text{abs}\in[0.75, 1.0]$ (b) and
		 $[0.5, 1.0]$ (d). The meaning of the colours is as in figure~\ref{fig:scheme3}.
		  The saw-tooth oscillations arise because we can use only discrete (integer) number of rounds 
		  with a kink appearing every time we change the number of rounds in our scheme
		  to stay close to the optimum.
    		}
      \label{fig:scheme3_scan}
 \end{figure*}

 We also investigated the susceptibility to photon loss by performing a scan of the success probability 
 and link fidelity for $p_\text{absorption}$ between 1 and 99\% and per-cycle loss $p_\text{loss}$ between 
 1 and 10\%.
 The results, shown in figure~\ref{fig:scheme3_scan}, show that while at least in approach B link 
 fidelity does not strongly depend on $p_\text{loss}$, the overall success probability quickly deteriorates 
 with increasing $p_\text{loss}$ confirming that it is indeed paramount to limit photon loss as far as possible.

 Figure~\ref{fig:scheme3_scan} also reveals a crossover between the two approaches A and B.
 While for low per-pass absorption $p_\text{abs}$, B yields superior fidelities, this advantage
 diminishes as $p_\text{abs}$ increases and almost disappears for $p_\text{abs}>90$\%, while
 success probability as defined here is always higher for scheme A. 
 Thus, given the final measurement can be implemented with high reliability, approach A has 
 a performance advantage in the high $p_\text{abs}$ regime, yielding an about $10$\% higher
 $p_\text{success}$ at comparable link fidelity (for more detailed numbers we refer to the 
 appendix). 
 
 \section{Relays and Repeaters}
 
Our heralded entanglement swapping operation is an extremely useful tool for the creation of long range 
entangled links. As was depicted in Figure (\ref{fig:teleport}), this tool can be used in a relay fashion to 
entangle a photon emitted from a remote NV center (say at Charlie location) with an already entangled link between two other 
remote NV centers (at Bob and Alice respective locations). Upon a successful entanglement swapping operation,
Alice and Charlie become entangled. David can then send a photon from his location to 
Charlie's location and the entanglement operation performed again. If it is successful, then Alice and David are entangled. 
In a relay fashion, longer range entanglement can be created. The probability of success for the creating this longer range 
links however will decrease exponentially with the numbers of nodes.  However using an entangled polarisation source 
of photons, two separate entangled links can be merged together (as is normally done in repeater networks). This in principle 
allows one to avoid this exponential issue. Further the entanglement swapping operations can be used 
to enable entanglement purification and so one has all the necessary elements for a quantum network.
 
\section{Conclusion}
  We have presented a proposal for a quantum communications node using only the electronic spin of
  a nitrogen vacancy center in diamond. It makes use of the Bell-type form of the NV's
  A$_2$ state to teleport the link carried by an incoming photon instantly when absorption
  is heralded by projective QND measurement. The quantum efficiency limit of 25\% is overcome 
  by periodically switching the photon's phase and/or polarisation, so that, eventually, all four Bell 
  states are detected.
  Modelling this scheme as a discrete-time quantum Markov process we were
  able to obtain a quantitative outline of the expected performance in the presence of some 
  real-world imperfections, foremost among which are photon loss and incomplete absorption
  per cycle. 
  The results of these simulations suggest, that photon loss per round should be reduce to below 
  0.5dB while at the same time increasing absorption to 50\% per pass or more. 
  If high absorption probabilities can be reached, an approach using a passive phase (polarisation)
  during each cycle with a final XX (ZZ)-basis measurement would offer the advantage of higher
  total success probability at comparable fidelities, provided high-efficiency
  single photon detectors are available.
  
\section{Acknowledgements}
 We thank Yuichiro Matsuzaki, Michael Hanks and Remi Blandino for valuable discussions. 
 This research was supported under the Commissioned Research of the National Institute of 
 Information and Communications Technology Quantum Repeater (A \& B) project and by a Grant-in-Aid 
 for Scientific Research (A)-JSPS (No. 24244044).
 
\appendix
\section{Parameter Estimation Continued}
 \label{app:paramestim}
 The starting state~\eqref{eq:psi0} contains four products of pairwise Bell-states btw. 1 and 3 
 as well as  2 and p respectively. 
 Only one, $\ket{\psi_-}_{2p}$, permits direct transition to $\ket{A_2}$, while the symmetry
 of the other three matches the other $m_S=\pm 1$ states in the ESM. Since these are split in 
 energy from $\ket{A_2}$ and thus detuned from the incoming photon, these transitions are 
 suppressed by a Lorentzian factor depending on
 detuning $\Delta\nu$ and the incoming photon's spectral width $\delta\nu$. The energetically
 closest state is $\ket{A_1}$ which is split by about 3GHz. Assuming the incoming photon is 
 spontaneously emitted by a remote NV, its lifetime is given by the $\ket{A_2}$ state's lifetime
 of approximately 10ns. From this the spectral width is $\delta\nu = 1/(10\pi\,\text{ns})\approx 30$MHz.
 Thus the transition to the nearest detuned state is already suppressed by a factor $1/(1+(\Delta\nu/\delta\nu)^2)\approx10^{-4}$.
 Transitions to $\ket{E_{1|2}}$ are roughly one order of magnitude smaller still which is why we
 chose not to include them in our model. 
 
 A similar argument can be made to assess the reliability of our QND measurement scheme.
 Here, there are actually two questions: given the NV center is in $A_2$, how certain are we
 to detect this, and if the system is not in $A_2$, but rather still in $m_S=\pm 1$ of the GSM, how 
 likely are we to see a false positive signal? In our model, the former is included in the form of 
 $p_\text{QND}$ while the latter is part of what we dubbed dark-count probability $p_\text{dark}$.
 As to the first case, the only way not to see a signal is that the system undergoes a transitions
 to either a dark state or another $m_S=\pm 1$ state. The first cannot occur for a perfect NV center
 but the possibility grows quadratically with applied strain and electric as well as magnetic fields. 
 All three influences can be reduced
 to the point where they are negligible: electric fields can be applied to cancel any remaining strain,
 external magnetic fields can be shielded and flip-flop processes with other spins in the sample,
 most notably nitrogen P1 centres, are suppressed by the energy splitting in the ESM.
 Transitions to another state could also be caused the driving field itself, in particular from the 
 starting state in the $m_S=\pm 1$ subspace to $m_S=0$. However, this can be avoided
 by choosing the Rabi power such that the Fourier transform of the driving field envelope
 is zero at this transition frequency ($2.88$GHz).
 We therefore choose a conservative $p_\text{QND}=0.99$ and $p_\text{dark} = 2\times 10^{-4}$.
 
 The dephasing times of the NV electronic spin at low temperature have been found to be
 as long as O(1ms)~\cite{} under the right conditions. 
 In our model we assume a commonly seen value of $100\mu$s and a total 
 cycle time of $\tau = \tau_\text{int} + \tau_\text{out} = 200$ns. The resulting dephasing
 error per NV involved is therefore $\gamma_2=1-\exp[-(0.005)^2]\approx 4\times 10^{-6}$. 
 Note that we assume the remote qubits 1 and 3 to be NV centres too, therefore they can be
 expected to experience the same kind of dephasing. So the worst case dephasing error is
 then roughly $3\gamma_2$ or $1.2\times 10^{-5}$ per cycle.
  
 The absorption probability is the hardest to model since it is very contingent on the sample
 and technical details of the implementation. Therefore we chose to use it as one of the
 'variable' parameters, along with the number of rounds $L$ (both approaches A and B)
 as well as the phase-and polarisation flip periods $l_x$ and $l_z$ (only approach B).

\section{Teleportation Errors}
 \label{app:errors}
 Here we look at the ways the teleportation scheme can fail within our model and estimate the 
 likelihood for the two types of error: false negatives and false positives.
 
 \subsection{False Negative}
 In this error scenario we do not see a click heralding teleportation, even though the photon got
 absorbed (and was subsequently spontaneously re-emitted and then lost) and the entanglement
 it was carrying was in fact teleported to the remote NV- center 3 (labels as in figure~\ref{fig:recycling}).
 This type of error will cause the whole scheme to fail, since in the absence of a heralding signal
 node NV2 will demand a new photon to be send from NV1, destroying the entangling link between
 1 and 3.
 Using our discrete model but neglecting (true) dark counts we find an upper bound for the probability 
 of this event
 \begin{equation}
  \begin{aligned}
   P_\text{false neg.}^\text{max} &= p_\text{abs} q_\text{QND} \sum_{l=0}^{L-1} q_\text{abs}^l p_\text{QND}^l \\
   						&=  p_\text{abs} q_\text{QND} \frac{1- (q_\text{abs} p_\text{QND})^{L} }{1- q_\text{abs} p_\text{QND}}
   \end{aligned}
    \label{eq:falsenegexact}
 \end{equation}
 with $q_\text{QND} \equiv 1-p_\text{QND}$ and $q_\text{abs} \equiv 1-p_\text{abs}$.
 The real value depend on the state $\rho_0$ but taking this into account can only reduce $p_\text{abs}$
 and thus lead to a lower total error.
 Using the result from the previous section, App.~\ref{app:paramestim}, we can set
 $p_\text{QND}\approx 1$ for the range of $L$ that are of interest. Thus~\eqref{eq:falsenegexact}
 simplifyies to $P_\text{false neg.}^\text{max} \approx q_\text{QND} (1-q_\text{abs}^{L}) \approx q_\text{QND}$.
 The latter approximation holds for the interesting regime $p_\text{abs} \gtrsim 0.5$.
 The results for the exact values of $P_\text{false neg.}$ for the different $p_\text{abs}$ 
 regimes investigated in the main text can be found in third column of table~\ref{tab:errors}.
 \begin{table}
   \begin{tabular}{|cc|cc|}
       \hline
       $p_\text{abs}$   &  	$L_\text{opt.}$	&  $P_\text{false neg.}^\text{max}/q_\text{QND}$   &   $P_\text{false pos.}^\text{max}/p_\text{dark}$ \\
       \hline
       1\%		&		40		&	$0.282$					    &	   27.5	 \\
       10\%		&		20		&	$0.836$					    &	    7.43	 \\
       25\%		&		20		&	$0.969$					    &      2.98       \\
       50\%		&		16		&	$0.990$					    &	    0.999      \\
       90\%		&		 4		&	$0.9988$					    &	    0.111	 \\
       \hline
   \end{tabular}
   \caption{Teleportation scheme errors: the false negative error causes the protocol to discard a teleported
   		link, but even in the interesting regime $p_\text{abs}\gtrsim 0.5$ it is never greater than $q_\text{QND}$
   		}
   \label{tab:errors}
 \end{table}
 
 \subsection{False Positive}
  A false positive error occurs when we see a click that was not actually cause by a correctly detected
  absorption event but either from a false detection event (proportional to $q_\text{QND}$ in our model)
  or a true dark count ($p_\text{Dark}$). The latter is a detection event independent of anything else. 
  The exact value again depends on the state $\rho$ but we can find an upper bound 
  in a similar manner as for false negatives:
 \begin{equation}
   \begin{aligned}
   P_\text{false pos.}^\text{max} &= p_\text{dark} q_\text{abs} \sum_{l=0}^{L-1} q_\text{abs}^l q_\text{dark}^l  \\
   						&=  p_\text{dark} q_\text{abs} \frac{1- (q_\text{abs} q_\text{dark})^{L} }{1- q_\text{abs} q_\text{dark}}
   \end{aligned}
    \label{eq:falseposexact}
 \end{equation}
 Making the approximation $q_\text{Dark}\approx 1$  and $q_\text{abs}^L\ll1$, this simplifies to 
 $P_\text{false neg.}^\text{max}\approx q_\text{abs} p_\text{dark}/p_\text{abs} $. Thus in the interesting
 absorption regime the overall influence of dark counts is in fact suppressed by the factor $q_\text{abs}/p_\text{abs} \lesssim 1$.
 The exact values are given in the fourth column of Table~\ref{tab:errors}.

\section{Detailed results}
Running simulations varying the number of rounds $L$ and the absorption probability per pas 
$p_\text{Abs}$ we found the success probabilities and fidelities given 
in Table~\ref{tabscheme3}.

\begin{widetext}
\begin{table*}
 \begin{center}
   \begin{tabular}{|c|cccc|cccccccc|}
   \hline
     				& \multicolumn{4}{c|}{approach A}		    			    	  		   	& \multicolumn{8}{c|}{approach B} \\
    $p_\text{absorp.}$&  $L$   &  $\;p_\text{success}\;$  &    F$_{\phi_\pm}$  &  F$_{\psi_\pm}$ 		& $l_z$  &  $l_x$  &  $L$   &  $\;p_\text{success}\;$  &  F$_{\phi_+}$  &  F$_{\phi_-}$  &  F$_{\psi_+}$  &  F$_{\psi_-}$   \\
    \hline		
     10\%		   	&  40      &  	43.3\%				&   0.97	     &		0.89	    		&   9       &  18      &    36   &   	33.0\%		&	0.993	&   0.991	     &		0.991      &	0.990  \\ 
      \hline	
     30\%		   	&  20      &     	61.6\%			      	&   0.975	     &		0.97    		&   6       &  12      &    24   &   	56.9\%		&	0.995     &   0.996	     &		0.995      &	0.996  \\	 
    \hline
      50\%		   	&   10      &        73.8\%			        &   0.98	     &		0.97   		&   4       &   8       &    16   &   	68.3\%		&	0.996        &   0.996	     &		0.996     &		0.997 \\
    \hline
    70\%		   	&   6       &      81.7\%			    	&   0.975	     &		0.97  		&   3       &   6       &    12   &   	76.0\%		&	0.996       &   0.996	     &		0.997     &		0.998  \\    
    \hline
     90\%		  	 &   4       &      86.9\%			   	&   0.985	     &		0.99 	 		&   2       &   4       &     8    &   	82.8\%		&	0.997       &   0.994	     &		0.993     &		0.995 \\	
   \hline
   \end{tabular}
  \end{center}
\caption{Single relay performance (success probability and average link fidelities) of the two approaches 
for different absorption probabilities and choices of total number of rounds $L$. Photon-loss per cycle was assumed to be .3dB or 6.6\%.}
\label{tabscheme3}	
 \end{table*}
  \end{widetext}

\bibliographystyle{apsrev}

\begin{thebibliography}{20}
\expandafter\ifx\csname natexlab\endcsname\relax\def\natexlab#1{#1}\fi
\expandafter\ifx\csname bibnamefont\endcsname\relax
  \def\bibnamefont#1{#1}\fi
\expandafter\ifx\csname bibfnamefont\endcsname\relax
  \def\bibfnamefont#1{#1}\fi
\expandafter\ifx\csname citenamefont\endcsname\relax
  \def\citenamefont#1{#1}\fi
\expandafter\ifx\csname url\endcsname\relax
  \def\url#1{\texttt{#1}}\fi
\expandafter\ifx\csname urlprefix\endcsname\relax\def\urlprefix{URL }\fi
\providecommand{\bibinfo}[2]{#2}
\providecommand{\eprint}[2][]{\url{#2}}


\bibitem{BB84} C. H. Bennett and G. Brassard,  \emph{Proceedings of IEEE International Conference on Computers, Systems, and Signal Processing}, Bangalore, India, IEEE Press (New York) 175-179 (1984).
%
\bibitem{Bennett96} C.H. Bennett, G. Brassard, S. Popescu, B. Schumacher, J. Smolin and W. K.  Wootters, Purification of Noisy Entanglement and Faithful Teleportation via Noisy Channels, Phys. Rev. Lett. {\bf 76}, 722 - 726 (1996).
%
\bibitem{enk98} S. Enk, J.I. Cirac and P. Zoller, Photonic channels for quantum communication, Science {\bf 279}, 205-208 (1998).
%
\bibitem{Gisin2002}  N. Gisin, G. Ribordy, W.Tittel and H.  Zbinden, Quantum cryptography, Rev. Mod. Phys. {\bf 74}, 145-195 (2002).
%
\bibitem{Gisin2007} N. Gisin and R. Thew, Quantum Communication, Nature Photon {\bf 1}, 165 - 171 (2007).
%
\bibitem{Kimble2008}H. J. Kimble, The quantum internet. Nature 453, 1023Ð1030 (2008).
%
\bibitem[{\citenamefont{Vazirani and Vidick}(2014)}]{Vazirani-14prl140501}
\bibinfo{author}{\bibfnamefont{U.}~\bibnamefont{Vazirani}} \bibnamefont{and}
  \bibinfo{author}{\bibfnamefont{T.}~\bibnamefont{Vidick}},
  \bibinfo{journal}{Phys. Rev. Lett.} \textbf{\bibinfo{volume}{113}},
  \bibinfo{pages}{140501} (\bibinfo{year}{2014}).
%
\bibitem{cirac1999} J.I. Cirac, A. Ekert, S. Huelga and C. Macchiavello, Distributed quantum computation over noisy channels. Phys. Rev. A 59, 4249 (1999).
%
\bibitem{meter2009}  R. Van Meter, W. Munro, K. Nemoto and K. Itoh, Arithmetic on a distributed- memory quantum multicomputer, ACM Journal on Emerging Technologies in Computing Systems 3, 2 (2009).
%
\bibitem{Giovannetti2004} V.Giovannetti, S.Lloyd, and L.Maccone, Quantum-Enhanced Measurements: Beating the Standard Quantum Limit, Science 306, 1330 (2004).
%
\bibitem{bennett1993} C.H. Bennett, G. Brassard, C. Crepeau, R. Jozsa, A. Peres, and W.K. Wootters, Teleporting an unknown quantum state via dual classical and Einstein-Podolsky-Rosen channels, Phys. Rev. Lett. 70, 1895 (1993).
%
\bibitem{Briegel1998} H.J. Briegel, W. Du\"r, J.I. Cirac and P Zoller, Quantum repeaters: The role of imperfect local operations in quantum communication. Phys. Rev. Lett. 81, 5932-5935 (1998).
%
\bibitem{Sangouard2011} N. Sangouard, C. Simon, N. de Riedmatten and N. Gisin, Quantum repeaters based on atomic ensembles and linear optics, Rev. Mod. Phys. 83, 33-80 (2011).
%
\bibitem{dur98} 
W. D\"ur, H.-J. Briegel, J. I.  Cirac, J\& P. Zoller,  Quantum repeaters based on entanglement purification. \textit{Phys. Rev. A} {\bf 59}, 169-181 (1999).
%
\bibitem{duan01}
L. M. Duan, M.D. Lukin, J.I. Cirac, \& P. Zoller,  Long-distance quantum communication with atomic ensembles and linear optics. \textit{Nature} {\bf 414}, 413-418 (2001).
%
\bibitem{loock06}
P. Van Loock, T. D. Ladd, K. Sanaka, F. Yamaguchi, K. Nemoto, W. J. Munro\&  Y. Yamamoto, Hybrid quantum repeater using bright coherent light. \textit{Phys. Rev. Lett.} {\bf 96}, 240501 (2006).
%
\bibitem{chen07a}
B. Zhao, Z. B. Chen, Y. A. Chen, J. Schmiedmayer\& J. W. Pan, Robust creation of entanglement between remote memory qubits. \textit{Phys. Rev. Lett.} {\bf 98}, 240502 (2007).
%
\bibitem{sheng08}
Z. Yuan, Y. Chen, B. Zhao, S. Chen, J. Schmiedmayer \& J. W. Pan, Experimental demonstration of a BDCZ quantum repeater node. \textit{Nature} {\bf 454}, 1098-1101 (2008). 
%
\bibitem{goebel08}
A. M. Goebel, G. Wagenknecht, Q. Zhang, Y. Chen, K. Chen, J. Schmiedmayer \& J. W. Pan, Multistage Entanglement Swapping.  \textit{Phys. Rev. Lett.} {\bf 101}, 080403 (2008).
%
\bibitem{simon07}
C. Simon, H. de Riedmatten, M. Afzelius, N. Sangouard, H. Zbinden \& N. Gisin, Quantum Repeaters with Photon Pair Sources and Multimode Memories.  \textit{Phys. Rev. Lett} {\bf 98}, 190503 (2007).
%
\bibitem{tittel08}
W. Tittel,  M. Afzelius, T. Chaneli\'{e}re, R. L. Cone, S. Kr\"oll, S. A. Moiseev \& M. Sellars, Photon-echo quantum memory in solid state systems. \textit{Laser Photon. Rev.} {\bf 4}, 244 - 267 (2009).
%
\bibitem{sangouard09}
N. Sangouard, R. Dubessy\& C. Simon, Quantum repeaters based on single trapped ions.  \textit{Phys. Rev. A} {\bf 79}, 042340 (2009).
%
\bibitem{pan01}
J. W. Pan, S. Simon, C. Brukner \& A. Zeilinger, Entanglement purification for quantum communication. \textit{Nature} {\bf 410}, 1067-1070 (2001).
%
\bibitem{dur07} 
W. D\"ur \& H. J. Briegel, Entanglement purification and quantum error correction. \textit{Rep. Prog. Phys.} {\bf 70}, 1381-1424 (2007).
%
\bibitem{munro10}
W. J. Munro, K. A. Harrison, A. M. Stephens, S. J. Devitt \& K. Nemoto, From quantum multiplexing to high-performance quantum networking. \textit{Nature Photon.} {\bf 4} , 792-796 (2010). 
%
\bibitem{munro12}
W. J. Munro, A. M. Stephens, S. J. Devitt, K. A. Harrison and Kae Nemoto, Quantum communication without the necessity of quantum memories, \textit{Nature Photon.} {\bf 6}  777 - 781 (2012).
%
\bibitem{Childress2006}
L.~Childress, J.~M.~Taylor, A.~S.~S\o{}rensen, and M.~D.~Lukin. Fault-Tolerant Quantum Communication Based on Solid-State Photon Emitters. {\it Phys. Rev. Lett.} {\bf 96}, 070504 (2006).
%
\bibitem[{\citenamefont{Stephens et~al.}(2013)\citenamefont{Stephens, Huang,
  Nemoto, and Munro}}]{Stephens-13pra052333}
\bibinfo{author}{\bibfnamefont{A.~M.} \bibnamefont{Stephens}},
  \bibinfo{author}{\bibfnamefont{J.}~\bibnamefont{Huang}},
  \bibinfo{author}{\bibfnamefont{K.}~\bibnamefont{Nemoto}}, \bibnamefont{and}
  \bibinfo{author}{\bibfnamefont{W.~J.} \bibnamefont{Munro}},
  \bibinfo{journal}{Phys. Rev. A} \textbf{\bibinfo{volume}{87}},
  \bibinfo{pages}{052333} (\bibinfo{year}{2013}).
  %
\bibitem{Davies1976}
G.~Davies and M.~F.~Hamer. Optical Studies of the 1.945 eV Vibronic Band in Diamond. {\it Proc. R. Soc. Lond. A} {\bf 348}, 285 (1976).
%
\bibitem{Harley1984}
R.~T.~Harley, M.~J.~Henderson, and R.~M.~Macfarlane. Persistent spectral hole burning of colour centres in diamond. {\it J. Phys. C} {\bf 17}, L233 (1984).
%
\bibitem{VanOort1988}
E.~Van~Oort, N.~B.~Manson, and M.~Glasbeek. Optically detected spin coherence of the diamond NV centre in its triplet ground state. {\it J. Phys. C: Sol. Stat. Phys.} {\bf 21}, 4385 (1988).
%
\bibitem{Doherty2013}
M.~W.~Doherty, N.~B.~Manson, P.~Delaney, F.~Jelezko, J.~Wrachtrup, and L.~C.~Hollenberg. The nitrogen-vacancy colour centre in diamond. Physics Reports {\bf 128} (1), 1-45 (2013).
%
\bibitem{Childress2005}
L. Childress, J. M. Taylor, A. S. S¿rensen, and M. D. Lukin,  Fault-tolerant quantum repeaters with minimal physical resources and implementations based on single-photon emitters. {\it Phys. Rev. A} {\bf 79},  052330 (2005).
%

\bibitem{Dutt2007}
M.~V.~G. Dutt, L.~Childress, L.~Jiang, E.~Togan, J.~Maze, F.~Jelezko, A.~S.~Zibrov, P.~R.~Hemmer, and M.~D.~Lukin. Quantum Register Based on Individual Electronic and Nuclear Spin Qubits in Diamond. {\it Science} {\bf 316}, 1312-1316 (2007).
%
\bibitem{Jiang2007}
L.~Jiang, J.~M.~Taylor, A.~S.~S\o{}rensen, and M.~D.~Lukin. Distributed quantum computation based on small quantum registers. {\it Phys. Rev. A} {\bf 76}, 062323 (2007).
%
\bibitem{YJG12}
N.~Yao, L.~Jiang, A.~Gorshkov, P.~Maurer, G.~Giedke, J.~Cirac, and M.~Lukin. Scalable Architecture for a Room Temperature Solid-State QuantumÊInformation Processor. {\it Nature Comm.} {\bf 3}, 800 (2012).
%
\bibitem{Maurer2012}
P.~C.~Maurer, G.~Kucsko, C.~Latta, L.~Jiang, N.~Y.~Yao, S.~D.~Bennett, F.~Pastawsk, D.~Hunger, N.~Chisholm, M.~Markham, D.~J.~Twitchen, J.~I.~Cirac, and M.~D.~Lukin. Room-Temperature Quantum Bit Memory Exceeding One Second. {\it Science} {\bf 336}, 1283-1286 (2012).
%
\bibitem{Togan2010}
E.~Togan, Y.~Chu, A.~S.~Trifonov, L.~Jiang, J.~Maze, L.~Childress, M.~V.~G.~Dutt, A.~S.~S\o{}rensen, P.~R.~Hemmer,  A.~S.~Zibrov, and M.~D.~Lukin. Quantum entanglement between an optical photon and a solid-state spin qubit. {\it Nature} {\bf 466}, 730-734 (2010).
%
\bibitem{Jelezko2004}
F.~Jelezko, T.~Gaebel, I.~Popa, A.~Gruber, and J.~Wrachtrup. Observation of coherent oscillations in a single electron spin. {\it Phys. Rev. Lett.} {\bf 92}, 076401 (2004).
%
\bibitem[{\citenamefont{Kennedy et~al.}(2003)\citenamefont{Kennedy, Colton,
  Butler, Linares, and Doering}}]{Kennedy-03apl4190}
\bibinfo{author}{\bibfnamefont{T.~A.} \bibnamefont{Kennedy}},
  \bibinfo{author}{\bibfnamefont{J.~S.} \bibnamefont{Colton}},
  \bibinfo{author}{\bibfnamefont{J.~E.} \bibnamefont{Butler}},
  \bibinfo{author}{\bibfnamefont{R.~C.} \bibnamefont{Linares}},
  \bibnamefont{and} \bibinfo{author}{\bibfnamefont{P.~J.}
  \bibnamefont{Doering}}, \bibinfo{journal}{Apl. Phys. Lett.}
  \textbf{\bibinfo{volume}{83}}, \bibinfo{pages}{4190} (\bibinfo{year}{2003}).
%
\bibitem[{\citenamefont{Jelezko
  et~al.}(2004{\natexlab{b}})\citenamefont{Jelezko, Gaebel, Domhan, Popa,
  Gruber, and Wrachtrup}}]{Jelezko-04prl130501}
\bibinfo{author}{\bibfnamefont{F.}~\bibnamefont{Jelezko}},
  \bibinfo{author}{\bibfnamefont{T.}~\bibnamefont{Gaebel}},
  \bibinfo{author}{\bibfnamefont{M.}~\bibnamefont{Domhan}},
  \bibinfo{author}{\bibfnamefont{I.}~\bibnamefont{Popa}},
  \bibinfo{author}{\bibfnamefont{A.}~\bibnamefont{Gruber}}, \bibnamefont{and}
  \bibinfo{author}{\bibfnamefont{J.}~\bibnamefont{Wrachtrup}},
  \bibinfo{journal}{Phys. Rev. Lett.} \textbf{\bibinfo{volume}{93}},
  \bibinfo{pages}{130501} (\bibinfo{year}{2004}{\natexlab{b}}).
%
\bibitem[{\citenamefont{Harrison et~al.}(2004)\citenamefont{Harrison, Sellars,
  and Manson}}]{Harrison-04jlum245}
\bibinfo{author}{\bibfnamefont{J.}~\bibnamefont{Harrison}},
  \bibinfo{author}{\bibfnamefont{M.~J.} \bibnamefont{Sellars}},
  \bibnamefont{and} \bibinfo{author}{\bibfnamefont{N.~B.}
  \bibnamefont{Manson}}, \bibinfo{journal}{J.Lumin.}
  \textbf{\bibinfo{volume}{107}}, \bibinfo{pages}{245} (\bibinfo{year}{2004}).
%
\bibitem[{\citenamefont{Balasubramanian
  et~al.}(2009)\citenamefont{Balasubramanian, Neumann, Twitchen, Markham,
  Kolesov, Mizuochi, Isoya, Achard, Beck, Tissler et~al.}}]{Bala-09nm383}
\bibinfo{author}{\bibfnamefont{G.}~\bibnamefont{Balasubramanian}},
  \bibinfo{author}{\bibfnamefont{P.}~\bibnamefont{Neumann}},
  \bibinfo{author}{\bibfnamefont{D.}~\bibnamefont{Twitchen}},
  \bibinfo{author}{\bibfnamefont{M.}~\bibnamefont{Markham}},
  \bibinfo{author}{\bibfnamefont{R.}~\bibnamefont{Kolesov}},
  \bibinfo{author}{\bibfnamefont{N.}~\bibnamefont{Mizuochi}},
  \bibinfo{author}{\bibfnamefont{J.}~\bibnamefont{Isoya}},
  \bibinfo{author}{\bibfnamefont{J.}~\bibnamefont{Achard}},
  \bibinfo{author}{\bibfnamefont{J.}~\bibnamefont{Beck}},
  \bibinfo{author}{\bibfnamefont{J.}~\bibnamefont{Tissler}},
  \bibnamefont{et~al.}, \bibinfo{journal}{Nature Materials}
  \textbf{\bibinfo{volume}{8}}, \bibinfo{pages}{383} (\bibinfo{year}{2009}).
%
\bibitem[{\citenamefont{Fuchs et~al.}(2010)\citenamefont{Fuchs, Dobrovitski,
  Toyli, Heremans, Weis, Schenkel, and D.}}]{Fuchs-10np668}
\bibinfo{author}{\bibfnamefont{G.~D.} \bibnamefont{Fuchs}},
  \bibinfo{author}{\bibfnamefont{V.~V.} \bibnamefont{Dobrovitski}},
  \bibinfo{author}{\bibfnamefont{D.~M.} \bibnamefont{Toyli}},
  \bibinfo{author}{\bibfnamefont{F.~J.} \bibnamefont{Heremans}},
  \bibinfo{author}{\bibfnamefont{C.~D.} \bibnamefont{Weis}},
  \bibinfo{author}{\bibfnamefont{T.}~\bibnamefont{Schenkel}}, \bibnamefont{and}
  \bibinfo{author}{\bibfnamefont{A.~D.} \bibnamefont{D.}},
  \bibinfo{journal}{Nature Physics} \textbf{\bibinfo{volume}{6}},
  \bibinfo{pages}{668} (\bibinfo{year}{2010}).
%
\bibitem[{\citenamefont{Wang et~al.}(2012)\citenamefont{Wang, de~Lange,
  Rist\`e, Hanson, and Dobrovitski}}]{wang-12prb155204}
\bibinfo{author}{\bibfnamefont{Z.-H.} \bibnamefont{Wang}},
  \bibinfo{author}{\bibfnamefont{G.}~\bibnamefont{de~Lange}},
  \bibinfo{author}{\bibfnamefont{D.}~\bibnamefont{Rist\`e}},
  \bibinfo{author}{\bibfnamefont{R.}~\bibnamefont{Hanson}}, \bibnamefont{and}
  \bibinfo{author}{\bibfnamefont{V.~V.} \bibnamefont{Dobrovitski}},
  \bibinfo{journal}{Phys. Rev. B} \textbf{\bibinfo{volume}{85}},
  \bibinfo{pages}{155204} (\bibinfo{year}{2012}).
%
\bibitem[{\citenamefont{Neumann et~al.}(2008)\citenamefont{Neumann, Mizuochi,
  Rempp, Hemmer, Watanabe, Yamazaki, Jacques, Gaebel, Jelezko, and
  Wrachtrup}}]{Neumann-08s1326}
\bibinfo{author}{\bibfnamefont{P.}~\bibnamefont{Neumann}},
  \bibinfo{author}{\bibfnamefont{N.}~\bibnamefont{Mizuochi}},
  \bibinfo{author}{\bibfnamefont{F.}~\bibnamefont{Rempp}},
  \bibinfo{author}{\bibfnamefont{P.}~\bibnamefont{Hemmer}},
  \bibinfo{author}{\bibfnamefont{H.}~\bibnamefont{Watanabe}},
  \bibinfo{author}{\bibfnamefont{S.}~\bibnamefont{Yamazaki}},
  \bibinfo{author}{\bibfnamefont{V.}~\bibnamefont{Jacques}},
  \bibinfo{author}{\bibfnamefont{T.}~\bibnamefont{Gaebel}},
  \bibinfo{author}{\bibfnamefont{F.}~\bibnamefont{Jelezko}}, \bibnamefont{and}
  \bibinfo{author}{\bibfnamefont{J.}~\bibnamefont{Wrachtrup}},
  \bibinfo{journal}{Science} \textbf{\bibinfo{volume}{320}},
  \bibinfo{pages}{1326} (\bibinfo{year}{2008}).
%
\bibitem[{\citenamefont{Neumann et~al.}(2010)\citenamefont{Neumann, Beck,
  Steine, Rempp, Fedder, Hemmer, Wrachtrup, and Jelezko}}]{Neumann-10s542544}
\bibinfo{author}{\bibfnamefont{P.}~\bibnamefont{Neumann}},
  \bibinfo{author}{\bibfnamefont{J.}~\bibnamefont{Beck}},
  \bibinfo{author}{\bibfnamefont{M.}~\bibnamefont{Steine}},
  \bibinfo{author}{\bibfnamefont{F.}~\bibnamefont{Rempp}},
  \bibinfo{author}{\bibfnamefont{H.}~\bibnamefont{Fedder}},
  \bibinfo{author}{\bibfnamefont{P.}~\bibnamefont{Hemmer}},
  \bibinfo{author}{\bibfnamefont{J.}~\bibnamefont{Wrachtrup}},
  \bibnamefont{and} \bibinfo{author}{\bibfnamefont{F.}~\bibnamefont{Jelezko}},
  \bibinfo{journal}{Science} \textbf{\bibinfo{volume}{329}},
  \bibinfo{pages}{542544} (\bibinfo{year}{2010}).
%
\bibitem[{\citenamefont{Robledo et~al.}(2011)\citenamefont{Robledo, Childress,
  Bernien, Hensen, Alkemade, and Hanson}}]{Robledo-11n574}
\bibinfo{author}{\bibfnamefont{L.}~\bibnamefont{Robledo}},
  \bibinfo{author}{\bibfnamefont{L.}~\bibnamefont{Childress}},
  \bibinfo{author}{\bibfnamefont{H.}~\bibnamefont{Bernien}},
  \bibinfo{author}{\bibfnamefont{B.}~\bibnamefont{Hensen}},
  \bibinfo{author}{\bibfnamefont{P.~F.} \bibnamefont{Alkemade}},
  \bibnamefont{and} \bibinfo{author}{\bibfnamefont{R.}~\bibnamefont{Hanson}},
  \bibinfo{journal}{Nature} \textbf{\bibinfo{volume}{477}},
  \bibinfo{pages}{574} (\bibinfo{year}{2011}).
%
\bibitem[{\citenamefont{Waldherr et~al.}(2014)\citenamefont{Waldherr, Wang,
  Zaiser, Jamali, Schulte-Herbrggen, Abe, Oshima, Isoya, Neumann, and
  Wrachtrup}}]{Waldherr-13n204}
\bibinfo{author}{\bibfnamefont{G.}~\bibnamefont{Waldherr}},
  \bibinfo{author}{\bibfnamefont{Y.}~\bibnamefont{Wang}},
  \bibinfo{author}{\bibfnamefont{S.}~\bibnamefont{Zaiser}},
  \bibinfo{author}{\bibfnamefont{M.}~\bibnamefont{Jamali}},
  \bibinfo{author}{\bibfnamefont{T.}~\bibnamefont{Schulte-Herbrggen}},
  \bibinfo{author}{\bibfnamefont{H.}~\bibnamefont{Abe}},
  \bibinfo{author}{\bibfnamefont{T.}~\bibnamefont{Oshima}},
  \bibinfo{author}{\bibfnamefont{J.}~\bibnamefont{Isoya}},
  \bibinfo{author}{\bibfnamefont{P.}~\bibnamefont{Neumann}}, \bibnamefont{and}
  \bibinfo{author}{\bibfnamefont{J.}~\bibnamefont{Wrachtrup}},
  \bibinfo{journal}{Nature} \textbf{\bibinfo{volume}{506}},
  \bibinfo{pages}{204} (\bibinfo{year}{2014}).
%
\bibitem[{\citenamefont{Taminiau et~al.}(2014)\citenamefont{Taminiau, Cramer,
  van~der Sar, Dobrovitski, and Hanson}}]{Taminiau-13nn171}
\bibinfo{author}{\bibfnamefont{T.}~\bibnamefont{Taminiau}},
  \bibinfo{author}{\bibfnamefont{J.}~\bibnamefont{Cramer}},
  \bibinfo{author}{\bibfnamefont{T.}~\bibnamefont{van~der Sar}},
  \bibinfo{author}{\bibfnamefont{V.}~\bibnamefont{Dobrovitski}},
  \bibnamefont{and} \bibinfo{author}{\bibfnamefont{R.}~\bibnamefont{Hanson}},
  \bibinfo{journal}{Nat. Nano.} \textbf{\bibinfo{volume}{9}},
  \bibinfo{pages}{171} (\bibinfo{year}{2014}).
%
\bibitem[{\citenamefont{Togan et~al.}(2010)\citenamefont{Togan, Chu, Trifonov,
  Jiang, Maze, Childress, Dutt, Soerensen, Hemmer, S. et~al.}}]{Togan-10n09256}
\bibinfo{author}{\bibfnamefont{E.}~\bibnamefont{Togan}},
  \bibinfo{author}{\bibfnamefont{Y.}~\bibnamefont{Chu}},
  \bibinfo{author}{\bibfnamefont{A.~S.} \bibnamefont{Trifonov}},
  \bibinfo{author}{\bibfnamefont{L.}~\bibnamefont{Jiang}},
  \bibinfo{author}{\bibfnamefont{J.}~\bibnamefont{Maze}},
  \bibinfo{author}{\bibfnamefont{L.}~\bibnamefont{Childress}},
  \bibinfo{author}{\bibfnamefont{M.~V.~G.} \bibnamefont{Dutt}},
  \bibinfo{author}{\bibfnamefont{A.~S.} \bibnamefont{Soerensen}},
  \bibinfo{author}{\bibfnamefont{P.~R.} \bibnamefont{Hemmer}},
  \bibinfo{author}{\bibfnamefont{Z.~A.} \bibnamefont{S.}},
  \bibnamefont{et~al.}, \bibinfo{journal}{Nature}
  \textbf{\bibinfo{volume}{466}}, \bibinfo{pages}{730} (\bibinfo{year}{2010}).
%
\bibitem[{\citenamefont{Bernien et~al.}(2013)\citenamefont{Bernien, Hensen,
  Pfaff, Koolstra, Block, Robledo, Taminiau, Markham, Twitchen, Childress
  et~al.}}]{Bernien-13n12016}
\bibinfo{author}{\bibfnamefont{H.}~\bibnamefont{Bernien}},
  \bibinfo{author}{\bibfnamefont{B.}~\bibnamefont{Hensen}},
  \bibinfo{author}{\bibfnamefont{W.}~\bibnamefont{Pfaff}},
  \bibinfo{author}{\bibfnamefont{G.}~\bibnamefont{Koolstra}},
  \bibinfo{author}{\bibfnamefont{M.~S.} \bibnamefont{Block}},
  \bibinfo{author}{\bibfnamefont{L.}~\bibnamefont{Robledo}},
  \bibinfo{author}{\bibfnamefont{T.~H.} \bibnamefont{Taminiau}},
  \bibinfo{author}{\bibfnamefont{M.}~\bibnamefont{Markham}},
  \bibinfo{author}{\bibfnamefont{D.~J.} \bibnamefont{Twitchen}},
  \bibinfo{author}{\bibfnamefont{L.}~\bibnamefont{Childress}},
  \bibnamefont{et~al.}, \bibinfo{journal}{Nature}
  \textbf{\bibinfo{volume}{497}}, \bibinfo{pages}{86} (\bibinfo{year}{2013}).
%
\bibitem[{\citenamefont{Kosaka and Niikura}(2015)}]{Kosaka-15prl053603}
\bibinfo{author}{\bibfnamefont{H.}~\bibnamefont{Kosaka}} \bibnamefont{and}
  \bibinfo{author}{\bibfnamefont{N.}~\bibnamefont{Niikura}},
  \bibinfo{journal}{Phys. Rev. Lett.} \textbf{\bibinfo{volume}{114}},
  \bibinfo{pages}{053603} (\bibinfo{year}{2015}).

\end{thebibliography}

\end{document}